\documentclass[preprint,12pt]{elsarticle}

\usepackage{graphicx,subfigure}
\usepackage{amssymb}
\usepackage{amsmath}
\usepackage{bm}
\usepackage{lineno}
\usepackage{natbib}
\journal{Computer Physics Communications}
\begin{document}

\begin{frontmatter}

\title{Numerical instability due to relativistic plasma drift in EM-PIC simulations}

\author[THUACC,THULPR]{Xinlu Xu}
\author[UCLAEE]{Peicheng Yu} 
\ead{tpc02@ucla.edu}
\author[IST]{Samual F. Martins}
\author[UCLAPH]{Frank S. Tsung}
\author[UCLAPH]{Viktor K. Decyk}
\author[IST]{Jorge Vieira}
\author[IST,ISCTE]{Ricardo A. Fonseca}
\author[THUACC,THULPR]{Wei Lu}
\author[IST]{Luis O. Silva}
\author[UCLAEE,UCLAPH]{Warren B. Mori}

\address[THUACC]{Department of Engineering Physics, Tsinghua University, Beijing 100084, China}
\address[THULPR]{Key Laboratory of Particle and Radiation Imaging of Ministry of Education, Tsinghua University, Beijing 100084, China}
\address[UCLAEE]{Department of Electrical Engineering, University of California Los Angeles, Los Angeles, CA 90095, USA}
\address[UCLAPH]{Department of Physics and Astronomy, University of California Los Angeles, Los Angeles, CA 90095, USA}
\address[IST]{Instituto Superior T\'ecnico, Lisbon, Portugal}
\address[ISCTE]{ISCTE - Instituto Universit\'ario de Lisboa, 1649--026, Lisbon, Portugal}

\begin{abstract}
The numerical instability observed in the Electromagnetic-Particle-in-cell (EM-PIC) simulations with a  plasma drifting with relativistic velocities is studied using both theory and computer simulations. We derive the numerical dispersion relation for a cold plasma drifting with a relativistic velocity and find an instability attributed to the coupling between the beam modes of the drifting plasma and the electromagnetic modes in the system. The characteristic pattern of the instability in Fourier space  for various simulation setups and Maxwell Equation solvers are explored by solving the corresponding numerical dispersion relations. Furthermore, based upon these characteristic patterns we derive an asymptotic expression for the instability growth rate. The asymptotic expression greatly speeds up the calculation of instability growth rate and makes the parameter scan for minimal growth rate feasible even for full three dimensions. The results are compared against simulation results and good agreement is found. These results can be used as a guide to develop possible approaches to mitigate the instability. We examine the use of a spectral solver and show that such a solver when combined with a low pass filter with a cutoff value of $\vert\vec{k}\vert$ essentially eliminates the instability while not modifying modes of physical interest. The use of spectral solver also provides minimal errors to electromagnetic modes in the lowest Brillouin zones.
\end{abstract}


\end{frontmatter}


\section{Introduction}
\label{sect:intro}

The effect of finite grid size and finite time step in electromagnetic simulations using the particle-in-cell method has been extensively studied both theoretically and in simulations \cite{Lindman1970Langdon1970,Birdsall1985,Godfrey1974,Godfrey1975}. Understanding these effects are crucial when one tries to separate numerical artifacts from real plasma phenomena. In recent years simulations with plasmas drifting with relativistic speeds have been conducted extensively, owing to using Lorentz boosted frames for the study of Laser Wakefield Acceleration (LWFA) \cite{Vay2007PRL,VayJCP2011,Martins2010NatPhys,MartinsCPC2010} and for the study of relativistic collisionless shocks. These simulations have revealed a  numerical instability which only occurs in multi-dimensions and which limits the range of parameters that can be explored in Lorentz boosted frame simulations of LWFA and relativistic shocks \cite{shock1,shock2}.  Godfrey studied the numerical instability induced by the a plasma drifting in 1D \cite{Godfrey1974} and in multi-dimensions \cite{Godfrey1975}. These analysis did not include relativistic mass effects in the plasma response (and therefore did not include relativistic drifts) and were applied to a code that solved the scalar and vector potential using the Coulomb gauge. However, most present day codes solve directly for the electric and magnetic fields  and use a rigorous charge conserving current deposition. Recent results indicate that there is only an instability in   2D and 3D  \cite{VayJCP2011,MartinsCPC2010}. 

To better understand and with the hope of mitigating  the observed instability, we present here an analysis of the numerical instability for a plasma drifting relativistically in multi-dimensions in which the electric field $E$ and magnetic field $B$ are solved for directly and where only a current deposit is included. We follow the basic method and notation of \cite{Godfrey1975} and concentrate on situations where the plasma is cold but  is drifting near the speed of light. The dispersion relation can be applied for various Maxwell field solvers and it includes finite size particle and aliasing effects. The dispersion relation reduces to purely longitudinal plasma waves and purely transverse light waves in a drifting plasma in the appropriate limits. The dispersion relation predicts the growth rates and range (pattern)  of unstable modes in  Fourier space. By comparing the theoretical predictions against  the simulation results using the EM-PIC FDTD code OSIRIS \cite{OSIRIS} and the UCLA PIC Framework \cite{UPIC} which is based on spectral solver, we conclude that the observed instability is indeed induced by the relativistic plasma drift (and not due to under-resolved backscattered radiation). It is found that the unstable modes lie near the intersection between   beam modes and transverse EM modes. We use this fact to  develop asymptotic expressions for the instability growth rate.  These observations can then be used as a guide for selecting alternative Maxwell  Equation solvers and smoothing schemes to mitigate the instability. Specifically, we show that a spectral solver together with a cutoff filter in $\vec k$ space can eliminate the instability.
 
The remainder of this paper is organized as follows. We first derive the multi-dimensional numerical dispersion relation in section \ref{sect:numdisp}. In section \ref{sect:numinst}, we use the 2D dispersion relation obtained in section \ref{sect:numdisp} to study the numerical instability induced by relativistic plasma drift. The theoretical results are compared against the simulation results, and good agreement is found. After including a minor correction in an earlier version our dispersion relation now predicts the time steps for which the minimal instability growth rate was observed after reading Ref. \cite{Godfreyarxiv}. We then use these results as a guide to discuss the methods for mitigating this instability. It is found that by using  a spectral solver to advance the EM field in Fourier space (which supports light waves with phase velocity greater than plasma drifting velocity), we can obtain desirable patterns of the instability which makes the mitigation of it more convenient. To better explore the instability in 3D scenario, we developed an asymptotic form of the instability growth rate in section \ref{sect:asym}, and used it to explain the observation that minimal growth rates occur under certain time steps, as reported in \cite{VayJCP2011}. Conclusions and direction for future work are presented in section \ref{sect:conclusion}. The detailed form of the field interpolation functions, as well as the finite difference operator used in this paper are listed in \ref{sect:app:s}. 


\section{Numerical dispersion relation for cold plasma drift}
\label{sect:numdisp}

\subsection{Derivation of dispersion relation}
\label{sect:numdisp:derivation}
We  mainly follow the notation in Ref. \cite{Godfrey1975} to derive the numerical dispersion relation for a cold plasma drifting with relativistic speeds. We note that the multi-dimensional analysis in \cite{Godfrey1975} solves for $\phi$ and $\vec A$ and is not valid for relativistic drifts, and the 1D analysis in \cite{Godfrey1974} predicts growth in 1D. On the other hand, our analysis includes relativistic mass effects, is valid in multi-dimensions, and it predicts instability only in multi-dimensions (in agreement with our simulations). Since most  EM-PIC codes now in use solve for  the electric field $\vec{E}$ and magnetic field $\vec{B}$ directly (with finite difference or spectral solvers), we derive a numerical dispersion relation  directly using these two quantities. Gaussian units will be used; in addition, particle mass and velocity will be normalized to electron mass and the speed of light.

For a multi-dimensional simulation setup in Cartesian coordinates, the EM field interpolated  on a particle can be expressed as 
\begin{align}
\vec{E}(t,\vec{x})&=\sum_{m,\vec{n}}\overleftrightarrow{S_E}(t,m,\vec{x},\vec{n})\vec{E}_{m,\vec{n}}\nonumber\\
\vec{B}(t,\vec{x})&=\sum_{m,\vec{n}}\overleftrightarrow{S_B}(t,m,\vec{x},\vec{n})\vec{B}_{m,\vec{n}}\nonumber
\end{align}
where $m$ is the time index and $\vec{n}$ is the grid index; $\overleftrightarrow{S}$ is the interpolation dyadic used to obtain the appropriate field at $\vec{x}$ and $t=m\Delta t$; $\vec{E}_{m,\vec{n}}$ and $\vec{B}_{m,\vec{n}}$ stands for the electromagnetic forces at time grid index $m$ and space grid index $\vec{n}$. For momentum conserving field interpolation $\overleftrightarrow{S_E}$ and $\overleftrightarrow{S_B}$ are equal and are scalar functions times the unit dyadic while for energy conserving field interpolation $\overleftrightarrow{S_E}$ and $\overleftrightarrow{S_B}$ are not equal in each direction (the interpolation dyadics for $\vec E$, $\vec B$, and $\vec j$ are given in \ref{sect:app:s}). The momentum change of the particle is related to the change in the distribution function of the plasma by the linearized Vlasov equation
\begin{align}
\frac{\partial }{\partial t}f(t,\vec{x},\vec{p})+\frac{\vec{p}}{\gamma}\cdot\frac{\partial }{\partial \vec{x}}f(t,\vec{x},\vec{p})+q\biggl\{\vec{E}(t,\vec{x})+\frac{\vec{p}}{\gamma}\times\vec{B}(t,\vec{x})\biggr\}\cdot\frac{\partial f_0}{\partial \vec{p}}=0\nonumber
\end{align}
where $\vec{p}$ is the particle momentum, and $\gamma$ is the particle Lorentz factor. After Fourier transforming, the Vlasov Equation becomes
\begin{align}\label{eqvlasovfs}
f(\omega,\vec{k},\vec{p})=-i  {q}  \biggl\{\overleftrightarrow{S_E}(\omega,\vec{k})\vec{E}(\omega,\vec{k})+ \frac{\vec{p}}{\gamma}\times\{\overleftrightarrow{S_B}(\omega,\vec{k})\vec{B}(\omega,\vec{k})\}\biggr\}\cdot\frac{\partial f_0}{\partial \vec{p}}(\omega-\vec{k}\cdot\frac{\vec{p}}{\gamma})^{-1} 
\end{align}

Note that $\vec{E}$ and $\vec{B}$ are defined at the discrete grid position and discrete time step, so its Fourier transform in $(w,k)$ is periodic, i.e.,
\begin{align}\label{eqfourierperiodic}
\vec{E}(\omega,\vec{k})=\vec{E}(\omega',\vec{k}')\qquad \vec{B}(\omega,\vec{k})=\vec{B}(\omega',\vec{k}')
\end{align}
where
\begin{align}\label{eqkgwg}
\omega'&=\omega+\mu\omega_g\qquad \omega_g=\frac{2\pi}{\Delta t}\qquad \mu = 0,\pm 1, \pm 2, \ldots\nonumber\\
{k}'_i&={k}_i+\nu_i{k}_{gi}\qquad {k}_{gi}=\frac{2\pi}{\Delta {x}_i}\qquad \nu_i = 0,\pm 1, \pm 2, \ldots
\end{align}
Note that when the EM field are staggered (such as on a Yee mesh), there is an addition $(-1)^{\sum_i\nu_i}$ term in $\vec{E}(\omega',\vec{k}')$ and $\vec{B}(\omega',\vec{k}')$, where $\hat{i}$ is summed over the directions for which the EM fields are half-grid offset \cite{Godfreyarxivnote}; we absorb these additional terms into $\overleftrightarrow{S_E}$ and $\overleftrightarrow{S_B}$ to keep Eq. (\ref{eqfourierperiodic}) correct (see \ref{sect:app:s}).

Replacing $(w,\vec{k})$ with $(w', \vec{k}')$ in Eq. (\ref{eqvlasovfs}), and using Eq. (\ref{eqkgwg}), we obtain
\begin{align}\label{eqdisfs}
f(\omega',\vec{k}',\vec{p})=-i  {q}  \biggl\{\overleftrightarrow{S_E}(\omega',\vec{k}')\vec{E}(\omega,\vec{k})+ \frac{\vec{p}}{\gamma}\times\{\overleftrightarrow{S_B}(\omega',\vec{k}')\vec{B}(\omega,\vec{k})\}\biggr\}\cdot\frac{\partial f_0}{\partial \vec{p}}(\omega'-\vec{k}'\cdot\frac{\vec{p}}{\gamma})^{-1} 
\end{align}

The current density $\vec{j}$ due to the movement of the particles can be expressed as
\begin{align}
\vec{j}(t,\vec{x})&=q\int \overleftrightarrow{S_j}(\vec{x'}-\vec{x})\frac{\vec{p}}{\gamma}f(\{m+1/2\}\Delta t, \vec{x}',\vec{p}) d\vec{x}'d\vec{p}\nonumber
\end{align}
where $\overleftrightarrow{S_j}(\vec{x}'-\vec{x})$ is the dyadic for the current deposit.  After Fourier transforming we obtain
\begin{align}\label{eqjfs}
\vec{j}(\omega,\vec{k})&=q\sum_{\mu,\vec\nu}(-1)^{\mu}\int \frac{\overleftrightarrow{S_j}(-\vec{k}')\vec{p}}{\gamma}f(\omega',\vec{k}',\vec{p})d\vec{p}
\end{align}

We can now proceed in the normal way to obtain a dispersion relation. We start from Faraday's and Ampere's Law,
\begin{align}
\nabla\times\vec{E}&=-\frac{\partial \vec{B}}{\partial t}\nonumber\\
\nabla\times\vec{B}&= \frac{\partial \vec{E}}{\partial t}+4\pi \vec{j}\nonumber
\end{align}
which upon Fourier transforming gives, 
\begin{align}
\label{eqfaraday}[\vec{k}]_{E}\times\vec{E}&=[\omega]\vec{B}\\
\label{eqampere}[\vec{k}]_{B}\times\vec{B}&=- [\omega]\vec{E}-4\pi i\vec{j}
\end{align}
$[k]_E$ and $[k]_B$ are the finite difference operator of the corresponding Maxwell solver schemes for $\vec{E}$ and $\vec{B}$ fields. We follow the notation in Ref.~\cite{Godfrey1975}, and use $[\cdot]$ exclusively to indicate the finite difference operator. Applying $[\vec{k}]_B\times$ to both sides of Eq. (\ref{eqfaraday}), we end up with the coupled wave equation for $\vec{E}$ and $\vec{j}$,
\begin{align}\label{eq:waveEj}
([\omega]^2-[\vec{k}]_E\cdot[\vec{k}]_B+[\vec{k}]_E[\vec{k}]_B)\vec{E} =-4\pi i[\omega] \vec{j}
\end{align}

Using Eq. (\ref{eqampere}) and (\ref{eqjfs}), we could obtain
\begin{align}\label{eqEEfourier}
 ([\omega]^2-[\vec{k}]_E\cdot[\vec{k}]_B+[\vec{k}]_E [\vec{k}]_B)\vec{E} =-4\pi iq\sum_{\mu,\vec\nu}(-1)^{\mu}[\omega] \int  \frac{\overleftrightarrow{S_j}(-\vec{k}')\vec{p}}{\gamma}f(\omega',\vec{k}',\vec{p})d\vec{p}
 \end{align}
and if we normalize the distribution function such that $f_0=n_0f^n_0$, use the definition of plasma frequency
\begin{align}
\omega^2_p={4\pi q^2n_0}
\end{align}
and use the expression for the distribution function in Eq. (\ref{eqdisfs}), we finally obtain
\begin{align}\label{eqEdispS}
&([\omega]^2-[\vec{k}]_E\cdot[\vec{k}]_B+[\vec{k}]_E[\vec{k}]_B)\vec{E}\nonumber\\
=~&-\omega^2_p \sum_{\mu,\vec{\nu}}(-1)^\mu\biggl\{\int \frac{\overleftrightarrow{S_j}(-\vec{k}')\vec{p}d\vec{p}}{\gamma\omega'-\vec{k}'\cdot\vec{p}} \biggl\{[\omega]\overleftrightarrow{S_E}(\omega',\vec{k}')\vec{E}+ \frac{\vec{p}}{\gamma}\times\{\overleftrightarrow{S_B}(\omega',\vec{k}')([\vec{k}]_E\times\vec{E})\} \biggr\}\cdot\frac{\partial f_0}{\partial \vec{p}}\biggr\}
\end{align}
which is a generalized dispersion relation for a plasma of finite size particles drifting on a grid. We note that the use of additional smoothers and filters can be incorporated into the dispersion relation by adding additional $S_{SM} (\vec k')$ terms outside the summation over Brillouin zones (essentially it multiplies the $\omega^2_p$ term).

\subsection{Elements of dispersion relation  tensor}
We next examine the dispersion relation in the limit of a cold plasma including the possibility that the drift is near the speed of light. 
 \subsubsection{3D case}
Note that $\overleftrightarrow{S}$ for the fields and the current has only three diagonal elements ${S}_{1}$, ${S}_{2}$, ${S}_{3}$ in each case. In 3D, we can expand Eq. (\ref{eqEdispS}) explicitly as
\begin{align}\label{eq:expand}
~&\begin{pmatrix}
([\omega]^2-[k]_{E1}[k]_{B1}-[k]_{E2}[k]_{B2}-[k]_{E3}[k]_{B3})E_1+[k]_{E1}[k]_{B1}E_1+[k]_{E1}[k]_{B2}E_2+[k]_{E1}[k]_{B3}E_3\\
([\omega]^2-[k]_{E1}[k]_{B1}-[k]_{E2}[k]_{B2}-[k]_{E3}[k]_{B3})E_2+[k]_{E2}[k]_{B1}E_1+[k]_{E2}[k]_{B2}E_2+[k]_{E2}[k]_{B3}E_3\\
([\omega]^2-[k]_{E1}[k]_{B1}-[k]_{E2}[k]_{B2}-[k]_{E3}[k]_{B3})E_3+[k]_{E3}[k]_{B1}E_1+[k]_{E3}[k]_{B2}E_2+[k]_{E3}[k]_{B3}E_3
\end{pmatrix}\nonumber\\
=~&-\omega^2_p \sum_{\mu,\vec{\nu}}(-1)^\mu\int \frac{dp_1dp_2dp_3}{\gamma(\gamma\omega'-k'_1p_1-k'_2p_2-k'_3p_3)}
\begin{pmatrix}
S_{j1}p_1\\
S_{j2}p_2\\
S_{j3}p_3
\end{pmatrix}
\begin{pmatrix}
\partial f^n_0/\partial p_1\\
\partial f^n_0/\partial p_2\\
\partial f^n_0/\partial p_3
\end{pmatrix}^T\cdot\nonumber\\
&\begin{pmatrix}
\gamma[\omega]S_{E1}E_1+p_2S_{B3}([k]_{E1}E_2-[k]_{E2}E_1)+p_3S_{B2}([k]_{E1}E_3-[k]_{E3}E_1)\\
\gamma[\omega]S_{E2}E_2+p_3S_{B1}([k]_{E2}E_3-[k]_{E3}E_2)+p_1S_{B3}([k]_{E2}E_1-[k]_{E1}E_2)\\
\gamma[\omega]S_{E3}E_3+p_1S_{B2}([k]_{E3}E_1-[k]_{E1}E_3)+p_2S_{B1}([k]_{E3}E_2-[k]_{E2}E_3)
\end{pmatrix}
\end{align}

This can be rewritten as 
\begin{align}
\overleftrightarrow{\epsilon}(\omega,k)\vec{E}=\begin{pmatrix}
\epsilon_{11}& \epsilon_{12} & \epsilon_{13}\\ 
\epsilon_{21} & \epsilon_{22} & \epsilon_{23}\\ 
\epsilon_{31} & \epsilon_{32} &\epsilon_{33} 
\end{pmatrix}\begin{pmatrix}
E_1\\ 
E_2\\ 
E_3
\end{pmatrix}=0
\end{align}
where we note that $\overleftrightarrow{\epsilon}$ is not the dielectric tensor. In addition, we are most interested in a cold plasma that is drifting. For such a case, the unperturbed  distribution function is given by
\begin{align}\label{eq:initdis}
f^n_0=\delta(p_1-p_0)\delta(p_2)\delta(p_3)
\end{align}
where $p_0=\gamma v_0$, and $v_0$ is the drifting velocity of the plasma. Substituting the above form for $f_0$,  Eq. (\ref{eq:initdis}), into Eq. (\ref{eq:expand}), and carrying out the integration we  obtain after some algebra all the elements in the tensor as
\begin{align}\label{eqepsilon3d}
\epsilon_{11}&=[\omega]^2-[k]_{E2}[k]_{B2}-[k]_{E3}[k]_{B3}-\frac{\omega^2_p}{\gamma}\sum_{\mu,\vec{\nu}}(-1)^{\mu}\frac{S_{j1}\{S_{E1}[\omega]\omega' / \gamma^2+v^2_0(S_{B3}k'_2[k]_{E2}+S_{B2}k'_3[k]_{E3})\}}{(\omega'-k'_1v_0)^2}\nonumber\\
\epsilon_{12}&= [k]_{E1}[k]_{B2} - \frac{\omega^2_p}{\gamma}\sum_{\mu,\vec{\nu}}(-1)^{\mu}\frac{S_{j1}k'_2v_0(S_{E2}[\omega]-v_0S_{B3}[k]_{E1})}{(\omega'-k'_1v_0)^2}\nonumber\\
\epsilon_{13}&=[k]_{E1}[k]_{B3}- \frac{\omega^2_p}{\gamma}\sum_{\mu,\vec{\nu}}(-1)^{\mu}\frac{S_{j1}v_0k'_3(S_{E3}[\omega]-v_0S_{B2}[k]_{E1})}{(\omega'-k'_1v_0)^2}\nonumber\\
\epsilon_{21}&=[k]_{E2}[k]_{B1}-\frac{\omega^2_p}{\gamma}\sum_{\mu,\vec{\nu}}(-1)^{\mu}\frac{v_0S_{j2}S_{B3}[k]_{E2}}{\omega'-k'_1v_0}\nonumber\\
\epsilon_{22}&=[\omega]^2-[k]_{E1}[k]_{B1}-[k]_{E3}[k]_{B3}-\frac{\omega^2_p}{\gamma}\sum_{\mu,\vec{\nu}}(-1)^{\mu}\frac{S_{j2}(S_{E2}[\omega]-v_0S_{B3}[k]_{E1})}{\omega'-k'_1v_0}\nonumber\\
\epsilon_{23}&=[k]_{E2}[k]_{B3}\nonumber\\
\epsilon_{31}&=[k]_{E3}[k]_{B1}-\frac{\omega^2_p}{\gamma}\sum_{\mu,\vec{\nu}}(-1)^{\mu}\frac{v_0S_{j3}S_{B2}[k]_{E3}}{\omega'-k'_1v_0}\nonumber\\
\epsilon_{32}&=[k]_{E3}[k]_{B2}\nonumber\\
\epsilon_{33}&=[\omega]^2-[k]_{E1}[k]_{B1}-[k]_{E2}[k]_{B2}-\frac{\omega^2_p}{\gamma}\sum_{\mu,\vec{\nu}}(-1)^{\mu}\frac{S_{j3}(S_{E3}[\omega]-v_0S_{B2}[k]_{E1})}{\omega'-k'_1v_0}
\end{align}

The dispersion relation is then finally obtained from the condition that 
 \begin{align}\label{eqdetepsilon}
\textrm{Det}(\overleftrightarrow{\epsilon})=0
\end{align}
which is valid in any number of dimensions.
\subsubsection{1D and 2D case}
Much can be learned from examining the 1D and 2D limits to the general dispersion relation. In 1D simulations all physical quantities only depend on one coordinate $x_1$, hence $[\vec{k}]$, $\vec{k}$, and $\vec{k}'$  only have the $\hat{1}$-component. It follows then that  the elements of the $\overleftrightarrow{\epsilon}$  are
\begin{align}\label{eqepsilon1d}
& \epsilon_{11}= [\omega]^2- \frac{\omega^2_p}{\gamma} \sum_{\mu,\nu}(-1)^\mu \frac{S_{j1}S_{E1}[\omega]\omega'/\gamma^2}{(\omega'-k'_1v_0)^2}\nonumber \\
&\epsilon_{22}=\epsilon_{33}=[\omega]^2-[k]_{E1}[k]_{B1} - {\frac{\omega^2_p}{\gamma}}\sum_{\mu,\nu} (-1)^\mu\frac{S_{j2}(S_{E2}[\omega]-S_{B3}[k]_{E1}v_0)}{\omega'-k'_1v_0}\nonumber\\
&\epsilon_{12}=\epsilon_{13}=\epsilon_{21}=\epsilon_{23}=\epsilon_{31}=\epsilon_{32}=0
\end{align}
Using Eq. (\ref{eqdetepsilon}),  the dispersion relation for the 1D case is three uncoupled modes,  
\begin{align}\label{eqdis1d}
\epsilon_{11}=0\qquad \epsilon_{22}=0
\end{align}
where each mode corresponds to separate components of the electric fields $E_1$, $E_2$, and $E_3$ respectively. Each of these modes is numerically stable as long as $\Delta t$ is sufficiently small. If we take the limit $\Delta t\rightarrow 0$, and $\Delta x\rightarrow 0$, then  Eq. (\ref{eqepsilon1d}) and (\ref{eqdis1d}) reduce to the dispersion relations in a real drifting plasma (which is completely stable).

Similarly, the elements of $\overleftrightarrow{\epsilon}$ in the 2D limit can be  written as
\begin{align}\label{eqepsilon2d}
\epsilon_{11}&= {[\omega]^2} -[k]_{E2}[k]_{B2}-{\frac{\omega^2_p}{\gamma}}\sum_{\mu,\vec\nu}  (-1)^\mu \frac { S_{j1}(S_{E1}\omega' [\omega]/\gamma^2+ S_{B3}[k]_{E2}k_2'v_0^2)} {(\omega ' - k_1' v_0)^2}\nonumber\\
\epsilon_{12} &=[k]_{E1}[k]_{B2}-{\frac{\omega^2_p}{\gamma}}\sum_{\mu,\vec\nu}  (-1)^\mu\frac{k_2'v_0S_{j1}(S_{E2}[\omega]- S_{B3}v_0[k]_{E1})}{(\omega'-k_1'v_0)^2}\nonumber\\
\epsilon_{21} &= [k]_{E2}[k]_{B1}-{\frac{\omega^2_p}{\gamma}}\sum_{\mu,\vec\nu} (-1)^\mu\frac{ S_{j2}S_{B3}[k]_{E2}v_0}{\omega'-k_1'v_0}\nonumber\\
\epsilon_{22}&= {[\omega]^2} -[k]_{E1}[k]_{B1}-{\frac{\omega^2_p}{\gamma}} \sum_{\mu,\vec\nu} (-1)^\mu\frac{S_{j2}(S_{E2}[\omega]- S_{B3}[k]_{E1}v_0) }{\omega' - k_1'v_0}\nonumber\\
\epsilon_{33}&={[\omega]^2} -[k]_{E1}[k]_{B1}-[k]_{E2}[k]_{B2}-{\frac{\omega^2_p}{\gamma}} \sum_{\mu,\vec\nu} (-1)^\mu\frac{S_{j3}(S_{E3}[\omega]- S_{B2}[k]_{E1}v_0)}{\omega' - k_1'v_0}\nonumber\\
\epsilon_{13}&=\epsilon_{23}=\epsilon_{31}=\epsilon_{32}=0
\end{align}
Using Eq. (\ref{eqdetepsilon}), we can obtain the dispersion relation for the 2D case 
\begin{align}\label{eqdis2d}
\epsilon_{11} \epsilon_{22} -\epsilon_{12} \epsilon_{21} =0\qquad \epsilon_{33}=0
\end{align}
Note that $E_3$ is de-coupled from the other two directions. 

The numerical features of a particular simulation setup can now be investigated by solving the corresponding numerical dispersion relation. Due to the use of  the finite space and time steps, these dispersion relations not only contain terms from the lowest order Brillouin zones ($\mu=0$ and $\vec\nu=\vec 0$), but also the space aliasing (summation over $\vec\nu$) and time aliasing (summation over $\mu$) terms \cite{Lindman1970Langdon1970}. The elements of the interpolation dyadic $\overleftrightarrow{S}$, and finite difference operators $[\cdot]$ comes into the expression due to the finite difference treatments when depositing currents and EM fields, and when solving the Maxwell Equations. The modifications to the dispersion relation leads to numerical instability in the otherwise stable physical system \cite{Lindman1970Langdon1970,Birdsall1985,Godfrey1974,Godfrey1975}.

\subsection{Beam modes and EM modes}
The 1D, 2D, and 3D dispersion relations show that a drifting plasma leads to beam modes  in the dispersion relation that are associated with longitudinal oscillations. A beam mode roughly satisfies the dispersion relation  
\begin{align} \label{eq:beam mode}
(\omega'-k'_1v_0)^2=\frac{\omega_p^2}{\gamma^3} \sim 0
\end{align}

In addition, a drifting plasma also supports transverse EM waves that are described by the dispersion relation
\begin{align}\label{eq:waveEvac}
[\omega]^2-[\vec k]_E\cdot[\vec k]_B=\frac{\omega_p^2}{\gamma} \sim 0
\end{align}
which for the high gammas are the dispersion relation on the grid for transverse EM modes in vacuum. In 3D, as the three components of the electric field are coupled, we expect to find instabilities near the intersections of the EM modes and beam modes in ($k_1$,  $k_2$, and $k_3$) space. However, in 2D $E_3$ is de-coupled from $E_1$ and $E_2$; thus the instability (coupling) can only occur in $k_1$ and $k_2$ space. 

In 1D, all the three components of the electric field are de-coupled so instability can only occur if either the longitudinal beam or transverse EM mode are numerically unstable themselves. We next concentrate on the 2D case since it is easier than the 3D but it still has the possibility of numerically unstable modes. 

As we show next, we observe in the simulations and in the solutions to our dispersion relation that in fact the unstable wave numbers and frequencies lie at the intersection of $\omega'-k'_1v_0=0$ (which we refer to as a vacuum beam mode) and the vacuum dispersion relation for EM waves. We will use this observation to derive asymptotic growth rates for the instability in section \ref{sect:asym}.


\section{Numerical instability induced by relativistic plasma drift}
\label{sect:numinst}

\subsection{Theoretical analysis of 2D dispersion relation}
\label{sect:numinst:theo}
Without loss of generality, we use the results in section \ref{sect:numdisp} to study the numerical instability induced by the relativistic plasma drift in a 2D system. According to the dispersion relation in 2D, we expect to observe instability in $E_2$ (and $B_3$). By calculating the maximum imaginary part of $\omega$ for real values of  ($k_1,k_2)$ for Eq. (\ref{eqdis2d}), we can obtain the characteristic pattern of the instability in Fourier space, as well as the growth rate of the instability. We can also plot the real part of $\omega$ for $k_1, k_2$. These results can be used later to compare with the simulation results. 

The dispersion relation is general and can be used to examine different choices in Maxwell Equation solvers, in differences between energy and momentum conserving field interpolation, in differences between charge conserving and direct current deposition schemes, and the use of smoothing and low pass filters. In this paper, we are emphasizing that our dispersion relation agrees well with the simulation results for cases studied, that we can predict the region of unstable modes by plotting where the beam and EM modes intersect in $\vec k$ and $\omega$ space; that we can obtain an asymptotic expression for the growth in 3D which agrees well with the simulation for various finite difference solvers (including the values of $\Delta t$ tht minimize the growth rate); and the advantages of using a spectral solver from the point of view of eliminating the instability and not attempting to carry out a comprehensive survey of all available choices listed above. 

We illustrate the instability using a  2D case with the standard Yee solver \cite{Yee}. We choose the grid parameters and time step that satisfies the Courant Condition \cite{Birdsall1985} to eliminate known numerical instability from the EM modes. We use the parameters in Tab. \ref{tab:simpara}, and substitute the finite difference operator for the Yee solver (see \ref{sect:app:s}) into the 2D dispersion relation. We assume linear (area) interpolation, momentum conserving field interpolation, and a charge conserving current deposition (see Appendix A).

\begin{figure}[th]
\begin{center}\includegraphics[width=1\textwidth]{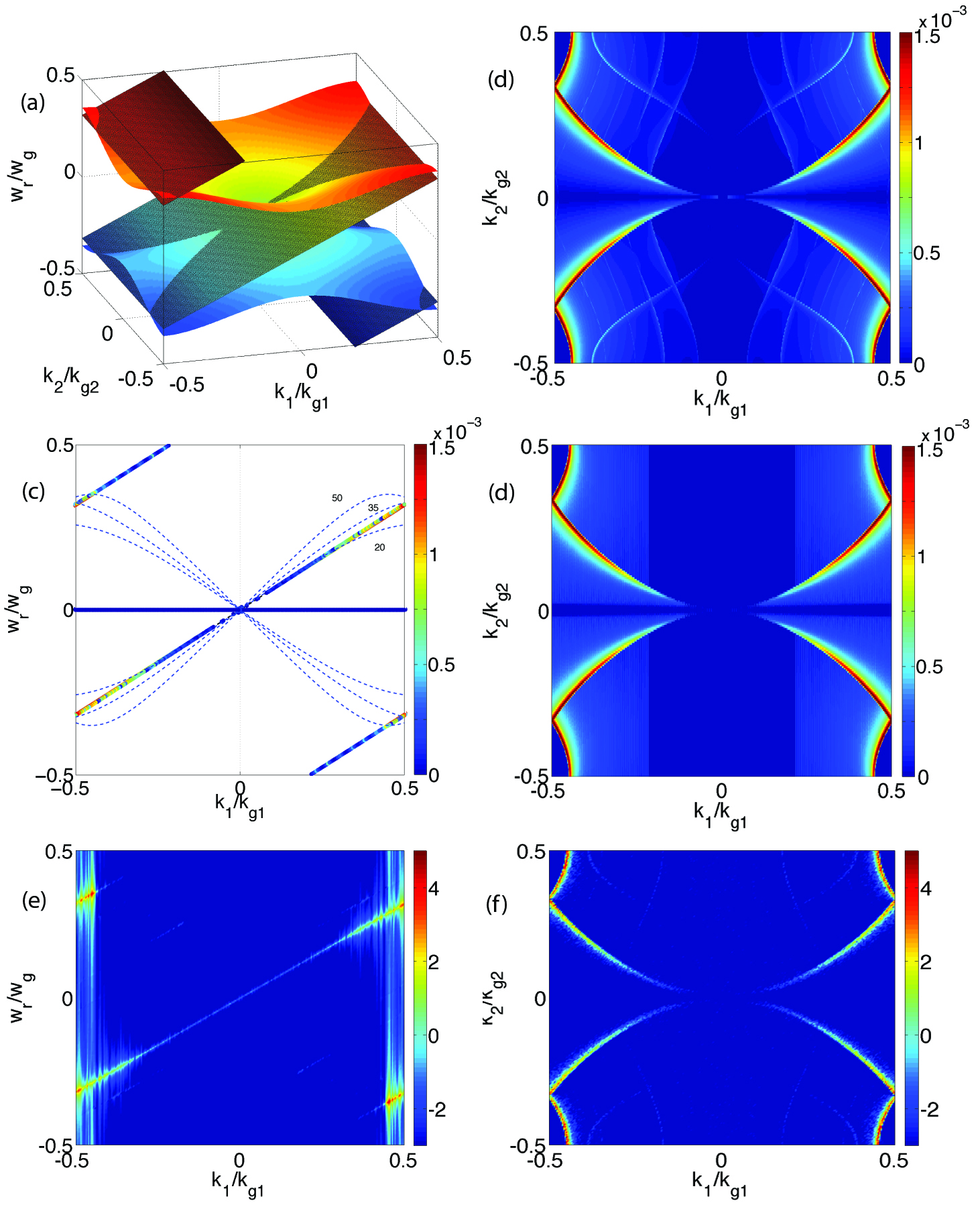}
\caption{\label{fig:yeeall} Numerical instability pattern in the Yee solver. Growth rate are color-coded, and normalized with $\omega_g$. (a) EM modes intersect with the main beam mode ($\mu=0$, $\nu=0$), and first order space aliasing modes ($\mu=0$, $\nu_1=\pm 1$); (b)  is the instability pattern ($\mu =0$, $\vert\nu_{1}\vert\le 4$) in $(k_1,k_2)$ space, plotted using Eq. (\ref{eqcubic2d2})--(\ref{eq:d2}); (c) and (d) are the instability pattern ($\vert\mu\vert \le 1$, $\vert\nu_{1}\vert\le 1$) in $(\omega_r, k_1)$ and $(k_1,k_2)$ spaces obtained from solving Eq. (\ref{eqepsilon2d}) and (\ref{eqdis2d}). EM modes for different propagating angles [in degree] and the beam modes are likewise plotted in (c). (e) presents the corresponding simulation results in $(\omega_r, k_1)$ space, and (f) in $(k_1,k_2)$ space. Data in (e) and (f) show the modes present  at $t=100~\omega^{-1}$, and are not a measurement of the growth rates.} 
\end{center}
\end{figure}

After obtaining all the roots $(\omega, k_1,k_2)$, we plot the dependence of the growth rate in the $(\omega_r,k_1)$ space [figure \ref{fig:yeeall} (c)], as well as in the $(k_1,k_2)$ space [figure \ref{fig:yeeall} (d)]. It is evident that all the instabilities are near the main or aliased beam modes. Since the terms with $\vert\mu\vert\le 1$, and $\vert\nu_{1}\vert\le 1$ are the most important, we neglect higher order terms when solving  Eq. (\ref{eqdis2d}). Higher order  $\mu$ and $\vec\nu$ terms can be included in the summation if needed. These additional terms lead to additional unstable modes in $(k_1, k_2)$ space  with lower growth rates as well as to small modifications to the growth rate and location of the original modes. A plot in $(k_1,k_2)$ space with more terms included are presented in figure \ref{fig:yeeall} (b), in which we use the asymptotic expression Eq. (\ref{eq:asym:2D}) for the growth rate (see section \ref{sect:asym} for more details). 

While the results in figure \ref{fig:yeeall} (c) and (d) are numerically calculated from Eq. (\ref{eqdis2d}), the location of the unstable modes  can also be conveniently predicted by plotting the intersection of the EM modes Eq. (\ref{eq:waveEvac}) and beam modes Eq. (\ref{eq:beam mode}) in $(k_1,k_2,\omega_r)$ space. This is shown in a 3D plot [figure \ref{fig:yeeall} (a)]. By examining the unstable pattern in  $(k_1,k_2)$ space we see that the central part of pattern comes from the intersections of the EM modes and main beam mode ($\mu=0$ and $\nu=0$), while the part at the four corners can be identified from the intersections of the EM modes and first order spatial aliasing beam modes ($\mu=0$ and $\nu_1=\pm 1$). As we argue in section \ref{sect:asym}, a key to mitigating the instability is to manipulate the instability pattern through a  careful choice of the Maxwell Equation solver. Making a  plot in $(k_1,k_2,\omega_r)$ of the intersection of the EM and beam modes for various solvers becomes a useful method for examining where the unstable modes reside without having to solve the full dispersion relation.

\begin{table}[t]
\centering
\begin{tabular}{p{8cm}c}
\hline\hline
\textbf{Parameters} & \textbf{Values}\\ 
\hline
solver & Yee\\
grid size $(k_0\Delta x_1, k_0\Delta x_2)$ & $(0.1,0.1)$\\
time step $\Delta t$ & $0.9\times$Courant limit\\
boundary condition & Periodic \\
simulation box size $(k_0L_1,k_0L_2)$ & 51.2$\times 25.6$\\
plasma drifting velocity & $\gamma=50.0$\\
plasma density & $n/n_0 = 1$\\
\hline\hline
\end{tabular}
\caption{Crucial simulation parameters for the 2D relativistic plasma drift simulation.}
\label{tab:simpara}
\end{table}

\subsection{Simulation study of the instability}
\label{sect:numinst:sim}

\begin{figure}[th]
\begin{center}\includegraphics[width=1.0\textwidth]{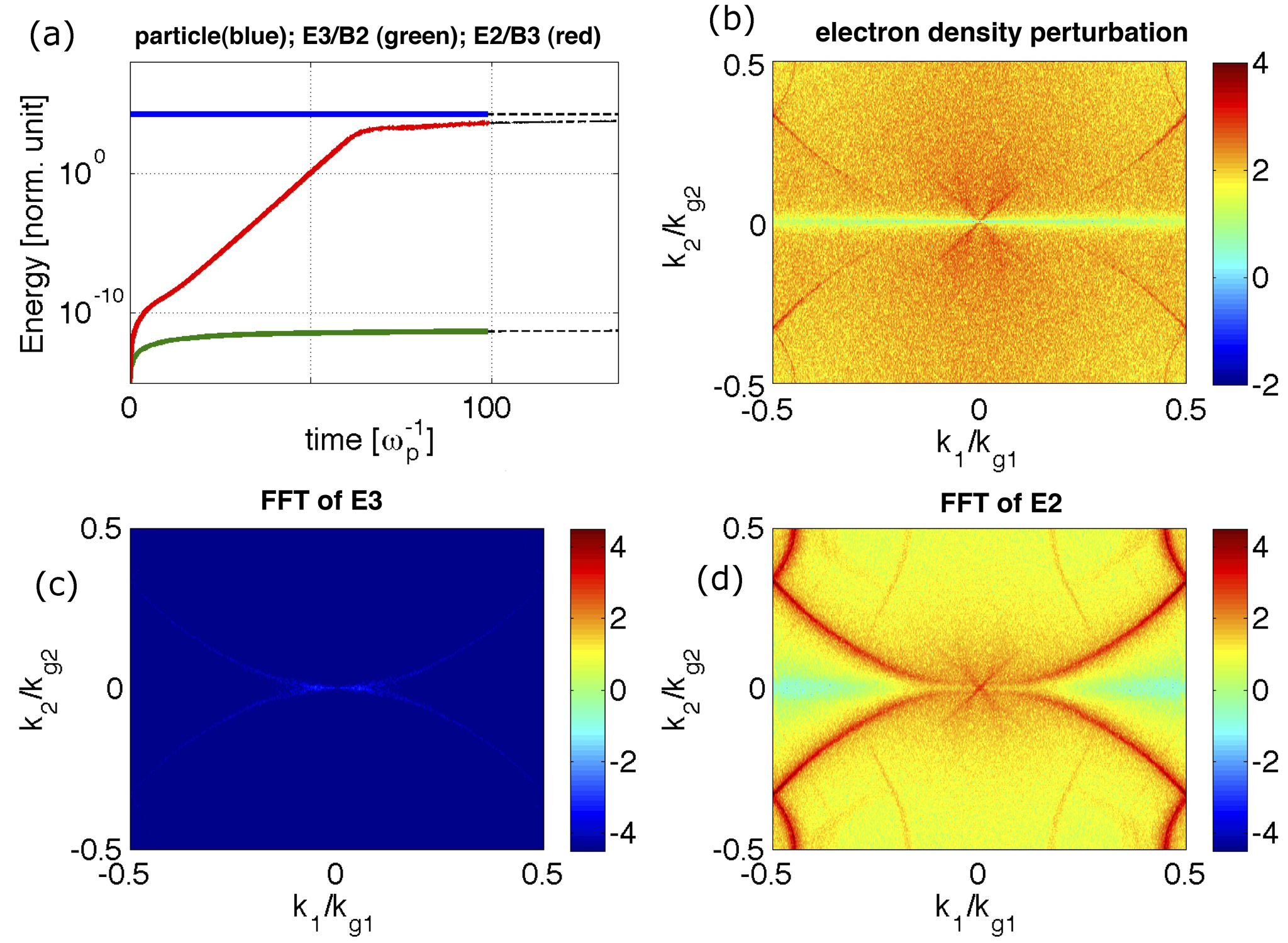}
\caption{\label{fig:yeesim} We present in (a) the energy evolution of the EM energy for the two cases. The corresponding dotted line indicates their variation in time after $t=100~\omega^{-1}_p$; (b) is the plasma electron density perturbation in $(k_1,k_2)$ space. (c) presents the $E_3$ in $(k_1,k_2)$ space, and (d) presents the $E_2$ in $(k_1,k_2)$ space.}
\end{center}
\end{figure}

To compare with the results in section \ref{sect:numinst:theo}, we conducted simulation studies in the 2D system using the EM-PIC code OSIRIS \cite{OSIRIS}. In these simulations, a neutral plasma with both the ion and electrons drifting in $x_1$ at the same relativistic velocity of $\gamma=50.0$ is initialized throughout the entire simulation box. Periodic boundary conditions for fields and particles are used. Other crucial parameters for the simulation setup are identical to the theoretical study in section \ref{sect:numinst:theo}. 

As is shown in figure \ref{fig:yeesim} (a), the total EM energy starts to grow violently as the plasma drifts relativistically. The exponential growth indicates that a numerical instability is occurring. In addition, the EM field energy in $E_2$ and $B_3$ and that in $E_3$ and $B_2$ are shown separately. As predicted by the 2D dispersion relation the $E_3$ and $B_2$ modes are stable and do not grow. The pattern of $E_2$ at $t=100~\omega^{-1}_p$ is plotted in figure \ref{fig:yeeall} (e) and (f), and good agreement for the location and relative amplitude of the unstable modes is obtained when compared against the theoretical prediction [figure \ref{fig:yeeall} (c) and (d)].

The EM energy grows with a lower rate after $t=110~\omega^{-1}_p$ [figure \ref{fig:yeesim} (a)]. The plasma density in this regime is highly modulated by the EM fields. The first order perturbation in plasma electron density [figure \ref{fig:yeesim} (b)] shows a  similar pattern as for $E_2$ [figure \ref{fig:yeesim} (d)] , which confirms they are coupling in the system. Note that no exponential energy growth can be seen in the $E_3$ field [figure \ref{fig:yeesim} (c)] 

From the simulation we find that for later times after the instability has evolved into a nonlinear state,  the same pattern in $(k_1,k_2)$ space as that of the linear regime still exists. This indicates that the instability will remain near the intersections of the EM modes and beam modes and that both the linear and nonlinear growth can be  mitigated through eliminating or controlling the intersections.

We also carried out a numerical investigation of the 1D dispersion relation Eq. (\ref{eqepsilon1d}), and (\ref{eqdis1d}) using the same simulation parameters as in Tab. \ref{tab:simpara} (with the 1D Courant condition). This confirms that there is no numerical instability under these parameters which is expected since  $E_1$ is de-coupled from $E_2$ and $E_3$ in Eq. (\ref{eqepsilon1d}) and each mode is itself stable.

We have done the same simulation studies on the use of other finite difference solvers beside the Yee solver, including the Karkkainen \cite{VayJCP2011,kark} and 4th-order solver \cite{4order}.  Good agreement between theory and simulations was also found \cite{YuAAC}. Some results are shown in figure \ref{fig:yeemcecasym} and discussed in section \ref{sect:asym:mts}.


\section{Mitigation of numerical instability}

\begin{figure}[th]
\begin{center}\includegraphics[width=1.0\textwidth]{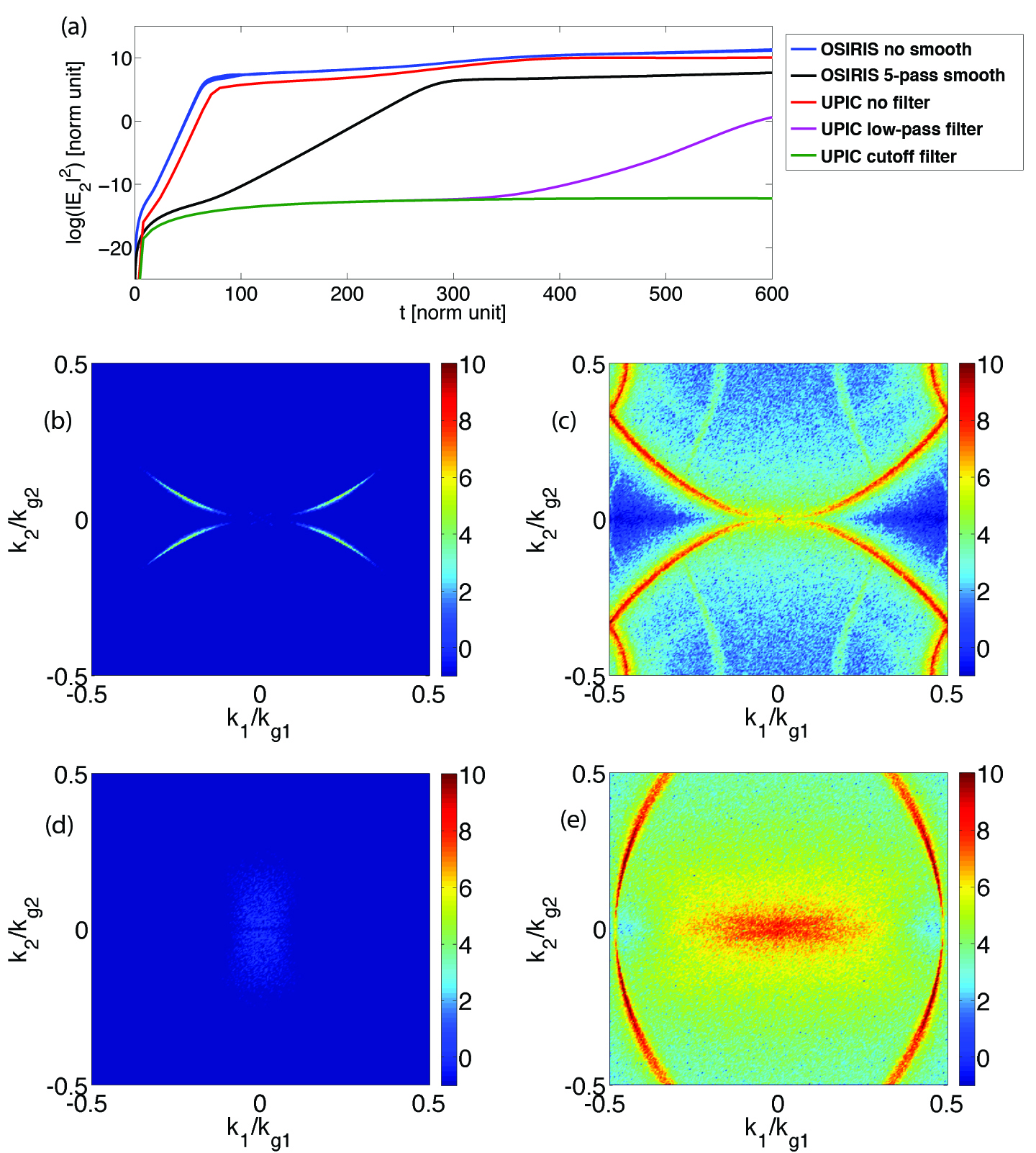}
\caption{\label{fig:yeesimsmooth} Instability mitigation in 2D simulation for the Yee and spectral solver. (a) Energy evolution of the $E_2$ field in various simulation setups; (b) presents $E_2$ in $(k_1, k_2)$ space for the Yee solver case in which a 4-pass binary smoothing with compensator is applied to the current, while (d) is the $E_2$ in $(k_1, k_2)$ space for spectral solver with cutoff smoothing. (c) and (e) is the $E_2$ field in $(k_1,k_2)$ space for the non-smoothing case for the Yee, and spectral solver respectively. (b)--(e) are plotted at $t=240~\omega^{-1}_p$. }
\end{center}
\end{figure}

\begin{figure}[th]
\begin{center}\includegraphics[width=1\textwidth]{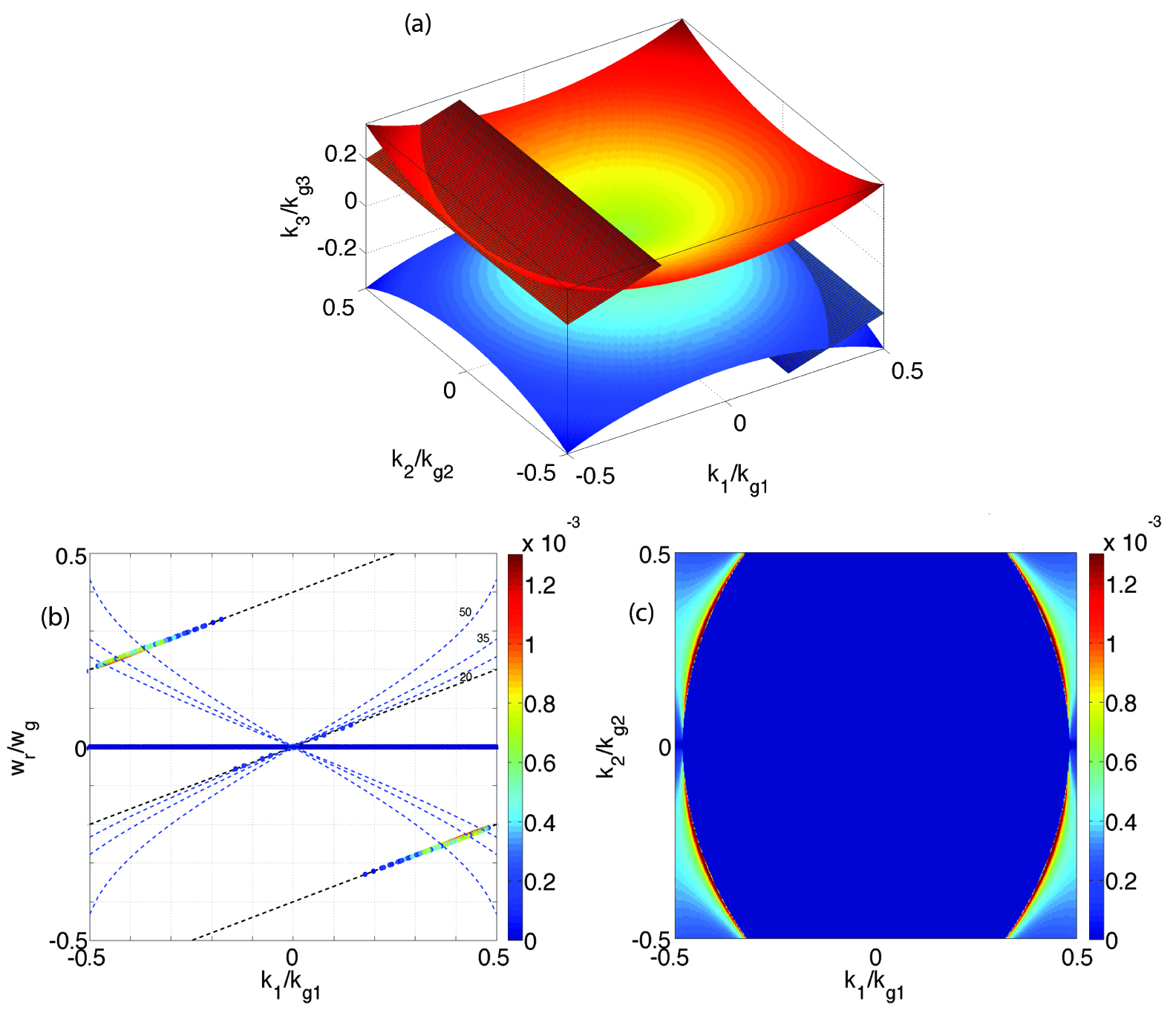}
\caption{\label{fig:spectralall} Numerical instability pattern for the spectral solver. Growth rate are color-coded, and normalized with $\omega_g$. (a) EM mode intersects with the first order space aliasing mode ($\mu=0$, $\nu_1=\pm 1$); (b) and (c) is the instability pattern in $(\omega_r, k_1)$, and $(k_1,k_2)$ spaces obtained from solving Eq. (\ref{eqepsilon2d}) and (\ref{eqdis2d}). Beam modes  ($\mu=0$, $\vert\nu_1\vert \le 1$) and EM modes for different propagating angles [in degree] are likewise plotted in (b).}
\end{center}
\end{figure}

Due to the fact that the EM dispersion curves for the Yee solver inevitably bends down at high $k_1/k_{g1}$ [c.f. figure \ref{fig:yeeall} (c)] such that the phase velocity of electromagnetic  (EM) waves on the grid is less than the plasma drifting velocity then an instability is found at the low $\vert\vec{k}\vert$ region in the $(k_1,k_2)$ space due to the fact that the beam mode can intersect the EM modes. In addition, there can be instability at high $\vert\vec{k}\vert$ near the intersection of the  EM mode  and the first order space aliasing beam mode ($\nu_{1}=\pm 1$). According to the numerical dispersion relation, the high $\vert\vec{k}\vert$ part due to aliasing can be conveniently smoothed out by applying low pass smoothing to the current (and/or EM field) when pushing the particles. However, the part due to intersection at the main beam mode is more difficult to mitigate, since the physics we want to simulate can reside in this region. 

Results from OSIRIS simulations with the Yee solver, linear order particles, and momentum conserving field interpolation with and without current smoothing are presented in figure \ref{fig:yeesimsmooth} (a)--(c). In the smoothing case 4-pass binary smoothing (1,2,1) with 1-pass compensator (-5,14,5) is applied to the current. We can see that the part near the first aliasing beam modes ($\nu_{1}=\pm 1$)  is greatly reduced, while that at the main beam mode ($\nu=0$) in the low $\vert\vec{k}\vert$ region is still present.

We next explore the use of a spectral solver to eliminate the intersection of the EM mode with the main beam mode. The use of  spectral solver to advance the EM field in Fourier space leads to the EM modes on the grid having phase velocities greater than the drifting velocity of the plasma which prevents any coupling with the main beam mode \cite{Godfrey1974,Nagata2007}.  As shown in figure \ref{fig:spectralall} (a), the intersection of the EM and beam modes occur first at the $\nu_1=\pm 1$ beam mode. Since the EM dispersion surface is a cone in $(k_1,k_2,\omega_r)$ plot, we would expect its intersection with the aliasing beam modes to be part of an ellipse that resides at high $\vert\vec{k}\vert$ region. Numerical results obtained by solving the corresponding numerical dispersion relation (spectral solver, linear shaped particles, momentum conserving field interpolation, and direct current deposit) are presented in figure \ref{fig:spectralall} (b) and (c), which confirms the empirical prediction in figure \ref{fig:spectralall} (a). 

Moreover, since we are  advancing the EM field in the Fourier space, it is easy to apply customized filters directly to the EM field. In figure \ref{fig:yeesimsmooth} (a), (d),  (e), we present simulation results using the spectral solver in the UCLA PIC Framework \cite{UPIC}. We show results where no filter is used, where a Gaussian shaped low-pass filter is used, and where a low pass filter with a hard cutoff is used. The Gaussian shaped filter was of the form,  $\exp(-\vert 
\vec{k}\vert^2/a^2)$, with $a=0.9k_1$ the RMS width of the filter in Fourier space, and the low pass filter with a hard cutoff set modes with $\vert \vec{k} \vert >0.9 k_1$ to zero. As we can see in the hard cutoff  case, the instability at $\nu_1=\pm 1$ is essentially eliminated from the simulation, thus only leaving instability with higher order terms.  This is confirmed by the fact that the  energy of the transverse EM modes remains at a very low level [figure \ref{fig:yeesimsmooth} (a)].


\section{Asymptotic expression for instability growth rate}
\label{sect:asym}

\subsection{Derivation of asymptotic expression}
In section \ref{sect:numinst}, we obtained the instability pattern and growth rate by numerically solving the dispersion relation equation, which is feasible on a modern laptop computer in 2D scenario, but much more difficult in 3D. As mentioned above, we observed the highest growth rate at the intersections of the beam modes and EM modes. Taking advantage of this observation, we here derive an asymptotic expression for the solutions near the beam mode $\omega' = k'_1 v_0$. These expressions will not only speed up the instability pattern analysis in 3D, but also provide more insights into the dependence of instability pattern and growth rate to the grid sizes and time step used in simulation. 

We denote the terms which are summing over $\mu$ and $\vec{\nu}$ in $\epsilon_{ij}$ as $Q_{ij}$, i.e.
\begin{align}
\epsilon_{11}&=[\omega]^2-[k]_{E2}[k]_{B2}-[k]_{E3}[k]_{B3}-Q_{11}\qquad
\epsilon_{12}= [k]_{E1}[k]_{B2} - Q_{12}\qquad
\epsilon_{13}=[k]_{E1}[k]_{B3}- Q_{13}\nonumber\\
\epsilon_{21}&=[k]_{E2}[k]_{B1}-Q_{21}\qquad
\epsilon_{22}=[\omega]^2-[k]_{E1}[k]_{B1}-[k]_{E3}[k]_{B3}-Q_{22}\qquad
\epsilon_{23}=[k]_{E2}[k]_{B3}\nonumber\\
\epsilon_{31}&=[k]_{E3}[k]_{B1}-Q_{31}\qquad
\epsilon_{32}=[k]_{E3}[k]_{B2}\qquad
\epsilon_{33}=[\omega]^2-[k]_{E1}[k]_{B1}-[k]_{E2}[k]_{B2}-Q_{33} \nonumber
\end{align}
We expand $\omega'$ around the beam mode $\omega' = k'_1v_0$, and write $\omega' = k'_1v_0+\delta\omega'$, where $\delta\omega'$ is a small term. In addition, we will use the relativistic limit $v_0\rightarrow 1$.  We will also truncate the summation over $\nu_2$ and $\nu_3$, keeping only the $\nu_2=\nu_3=0$ terms. Using $\det(\overleftrightarrow{\epsilon})=0$, and dropping terms of higher order of $(\omega^2/\gamma)^2$, $(\omega^2/\gamma)^3$, $\ldots$, we can obtain
\begin{align}\label{eqcubic2d}
A_1 \delta\omega'^2 +  B_1 \delta\omega'  +   C_1 =0   
\end{align}
where
\begin{align}
A_1&=[\omega]^2\left([\omega]^2-[k]_{E1}[k]_{B1}-[k]_{E2}[k]_{B2}-[k]_{E3}[k]_{B3}\right)  \nonumber\\
B_1&=[k]_{E1}[k]_{B2}Q_{21} + [k]_{E1}[k]_{B3}Q_{31} - \left( [\omega]^2-[k]_{E2}[k]_{B2} \right)Q_{22} - \left( [\omega]^2-[k]_{E3}[k]_{B3} \right) Q_{33}  \nonumber\\
C_1&= -\left( [\omega]^2 -[k]_{E1}[k]_{B1}\right) Q_{11} + [k]_{E2}[k]_{B1}Q_{12} +[k]_{E3}[k]_{B1} Q_{13}
\end{align}

Now we use the condition that $(\omega', k'_1)$ sits near the EM modes,
\begin{align}
[\omega]^2  \approx [k]_{E1}[k]_{B1}+[k]_{E2}[k]_{B2}+[k]_{E3}[k]_{B3}
\end{align}
We further expand the finite difference operator $[\omega]=\xi_0+\delta\omega' \xi_1$, where
\begin{align}
\xi_0=\frac{\sin(\tilde{k}\Delta t/2)}{\Delta t/2}\qquad \xi_1=\cos(\tilde{k}\Delta t/2)
\end{align}
and
\begin{align}
\tilde{k}=k_1+\nu_1k_{g1}-\mu\omega_g
\end{align}
We further expand $[w]$ to first order in $A_1$ since this term is sensitive near the EM mode, while keeping only zero order of $[w]$ in $B_1$, and $C_1$.  We then obtain a cubic equation
\begin{align}\label{eqcubic2d2}
A_2 \delta\omega'^3 +  B_2 \delta\omega'^2  +   C_2 \delta\omega'  + D_2=0   
\end{align}
with the coefficients 
\begin{align}
A_2&=2\xi_0^3 \xi_1  \nonumber\\
B_2&=\xi_0^2\left[ \xi_0^2 - \left( [k]_{E1}[k]_{B1} +[k]_{E2}[k]_{B2} +[k]_{E3}[k]_{B3}\right)\right]  \nonumber\\
C_2&=  [k]_{E1}[k]_{B2}Q_{21} + [k]_{E1}[k]_{B3}Q_{31} - \left( {\xi_0^2}-[k]_{E2}[k]_{B2} \right)Q_{22} - \left( {\xi_0^2}-[k]_{E3}[k]_{B3} \right) Q_{33}  \nonumber\\
\label{eq:d2}D_2&= - \left( \xi_0^2 -[k]_{E1}[k]_{B1}\right) Q_{11} + [k]_{E2}[k]_{B1}Q_{12} +[k]_{E3}[k]_{B1} Q_{13}
\end{align}
When calculating the instability growth rate, we obtain the imaginary part of roots $\Im\{\delta\omega'(\vec{k},\mu,\nu_1)\}$ for each $\mu$ and $\nu_1$ by solving Eq. (\ref{eqcubic2d2})--(\ref{eq:d2}), and the growth rate $\Gamma(\vec{k_0})$ for  a particular mode $\vec{k}_0$ is chosen to be $\max\{\Im\{\delta\omega'(\vec{k}_0,\mu,\nu_1)\}\}$. When solving for each $\Im\{\delta\omega'(\vec{k},\mu,\nu_1)\}$, we only keep the corresponding $\mu$ and $\nu_1$ terms in the above cubic equation. Eq. (\ref{eqcubic2d2})--(\ref{eq:d2}) can be used to plot the growth rate in Fourier space, and can be conveniently simplified to 2D. In figure \ref{fig:yeeall} (b) we plot the asymptotic instability growth rate with $\mu=0$, $\vert\nu_1\vert \le 4$ for 2D Yee solver using the same parameter listed in table \ref{tab:simpara}.

Near the transverse EM modes $[\omega]^2\approx \xi_0^2\approx  [\vec{k}]_E\cdot[\vec{k}]_B$, we can drop the $B_2$ term in Eq. (\ref{eqcubic2d2}). According to our numerical results, we can further simplify the analytical expressions by dropping the small $C_2$ term. The asymptotic growth rate $\Gamma(\vec{k})$ for this mode corresponds to the maximum imaginary part for the three roots,
\begin{align}
\label{eq:asym:2D}
\Gamma(\vec{k}) &\approx \frac{\sqrt{3}}{2}\left| \frac{ \left( \xi_0^2 -[k]_{E1}[k]_{B1}\right) Q_{11} - [k]_{E2}[k]_{B1}Q_{12} - [k]_{E3}[k]_{B1} Q_{13}}{2\xi_0^3\xi_1}\right|^{1/3} \nonumber \\
&\approx \frac{\sqrt{3}}{2}\left| \frac{ \omega_p^2 S_{j1} \left\{ (S_{B3}\xi_0 - S_{E2}[k]_{B1}) [k]_{E2}k_2 + (S_{B2}\xi_0 - S_{E3} [k]_{B1}) [k]_{E3} k_3\right\}}{2\gamma \xi_0^2\xi_1}\right|^{1/3}
\end{align}
This expression shows the relation between the instability growth rate and grid sizes, time step, interpolation and smoothing functions, and finite difference operators. Note that from the positions of the interpolation functions we could immediately see that using higher particle shape, or using a stronger smoother helps mitigate the instability pattern at high $\vert \vec{k}\vert$ region, which agrees well with our simulation.

\subsection{Parameter scans for minimal instability growth rate}
\label{sect:asym:mts}
With the asymptotic expression Eq. (\ref{eqcubic2d2})--(\ref{eq:d2}), and also Eq. (\ref{eq:asym:2D}), we can greatly speed up the solution of numerical dispersion relation in 2D and 3D. In addition, the asymptotic expression makes the parameter scan to study the dependence of instability pattern and growth rate between various grid sizes and time step more convenient. 

\begin{figure}[th]
\begin{center}\includegraphics[width=1\textwidth]{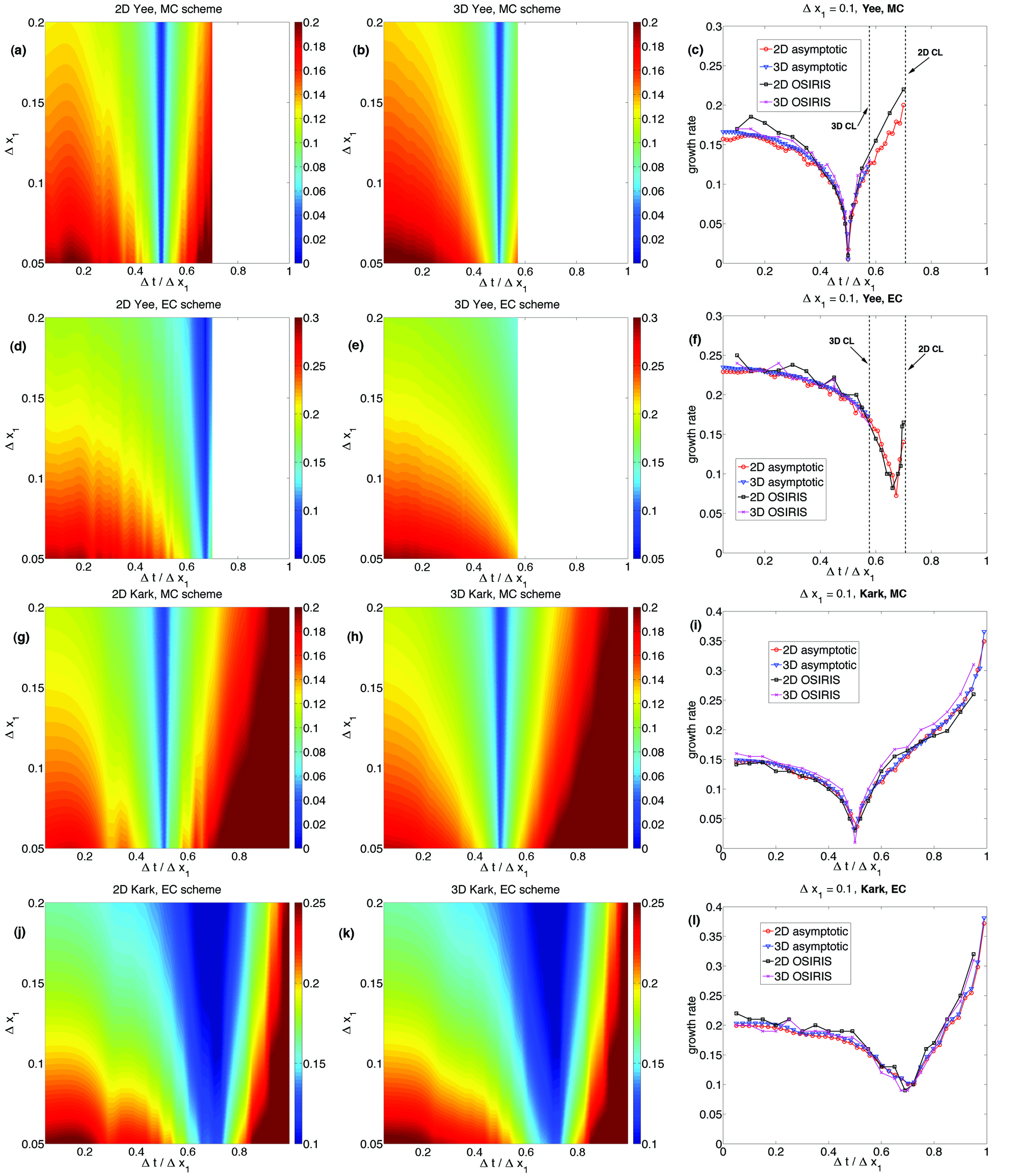}
\caption{\label{fig:yeemcecasym} Parameter scan of $\Delta x_1$ and $\Delta t/\Delta x_1$ for the Yee (first two rows), and Karkkainen (last two rows) solver. The first and third row uses momentum conserving (MC) scheme, while the second and fourth row uses energy conserving (EC) scheme. The simulation results are likewise plotted in (c), (f), (i), and (l) at $\Delta x_1=0.1$ for comparisons. In (c) and (f) the dotted line at $\Delta t/\Delta x_1 \approx 0.577$ is the 3D Courant limit (CL), and that at at $\Delta t/\Delta x_1 \approx 0.707$ is the 2D CL.}
\end{center}
\end{figure}

In figure \ref{fig:yeemcecasym}, we scanned the grid sizes $\Delta x_1$ and time step $\Delta t/\Delta x_1$ for the 2D and 3D Yee solver, and Karkkainen solver, and compared the growth rates with the OSIRIS simulations. We have kept $\Delta x_1=\Delta x_2(=\Delta x_3)$ during the parameter scan for 2D (and 3D). We likewise plotted out the OSIRIS simulation data for $\Delta x_1 = 0.1$ together with the asymptotic data (similar to the plots in Ref. \cite{Godfreyarxiv}). There are several interesting points worth noting in figure \ref{fig:yeemcecasym}. First, we can see there is a ``magic time step'' \cite{VayJCP2011} $\Delta t_m/\Delta x_1$ where the growth rate is minimized in most cases; on the other hand, the instability growth rate decreases monotonically as the grid sizes increases; second, when the grid sizes are square (2D) or cubic (3D), the ``magic time step'' $\Delta t_m/\Delta x_1$ is an invariant for different $\Delta x_1$, in both the momentum conserving (MC) scheme, and energy conserving (EC) scheme; third, the instability growth rate for 2D and 3D are nearly the same for given $\Delta x_1$ and $\Delta t/\Delta x_1$ under the same field interpolation scheme; the values for the magic time steps are also nearly the same in 2D and 3D (note that according to the asymptotic expression, the magic time step for Yee solver 3D EC scheme also resides at around  $\Delta t_m/\Delta x_1 \approx 0.65$, but we did not plot it out since that $\Delta t_m$ is beyond the Courant limit for this solver). The parameter scan using the asymptotic expression for the Karkkainen solver with EC scheme shows the ``magic time step'' at around $\Delta t_m/\Delta x_1 = 0.7$, which agrees with the results reported in Ref. \cite{VayJCP2011}. However, according to our simulation and theoretical results, we found the ``magic time step'' not only in Karkkainen solver, but also in Yee solver; and not only for EC scheme, but also for MC scheme. This is also reported in \cite{Godfreyarxiv} for the 2D cases.

The fact that the ``magic time step'' $\Delta t_m/\Delta x_1$ does not depend on $\Delta x_1$ for square (and cubic) cell for the Yee and Karkkainen solver is evident from Eq. (\ref{eq:asym:2D}). Applying the detailed form of finite difference operators $[k]_{Bi}$ for Yee and Karkkainen solver (note they have the same $[k]_{Bi}$, see \ref{sect:app:s}), for both MC and EC scheme, the expression of $\Gamma$ can be expressed as $\vert k'_1\vert^{1/3}$ times a function of $\Delta t/\Delta x_1$, and $k'_i/k_{gi}$. (this function is different for different field interpolation schemes). Since $k'_i/k_{gi}$ ranges from $(-0.5,0.5)$ regardless of $\Delta x_1$ when calculating the growth rate, the extreme value of $\Gamma$ resides at the same $\Delta t/\Delta x_1$ for different $\Delta x_1$ (although different field interpolation schemes give different ``magic time step''). In particular, in MC scheme the terms
\begin{align}
S_{B3}\xi_0-S_{E2}[k]_{B1}\qquad S_{B2}\xi_0-S_{E3}[k]_{B1}
\end{align}
(or only the first one in 2D) in Eq. (\ref{eq:asym:2D}) are zero when $\Delta t/\Delta x_1 = 1/2$ for these two solvers. As a result, both Yee and Karkkainen solver reach the minimal growth rate at  $\Delta t/\Delta x_1 = 1/2$ in MC scheme, which agrees well with OSIRIS simulations.

\section{Conclusions and future work}
\label{sect:conclusion}

We derived a general multi-dimensional numerical dispersion relation for the relativistic plasma drift in the EM-PIC simulation that can include different choices  in Maxwell solvers, differences between energy and momentum conserving field interpolation, differences between charge conserving and direct current deposition schemes, and the use of smoothing and low pass filters. In this paper we emphasized trying to understand the source of the instability and the structures of the dispersion relation. We confirmed that no instability occurs in 1D, and that in 2D and 3D a strong instability occurs due to the coupling between beam modes and transverse EM modes in the system. We can predict the pattern and growth rate of the instability for a particular simulation by solving the corresponding numerical dispersion relations. An asymptotic expression which permits rapid parameter scan for the ranges of unstable modes was derives. These results are compared against simulation results using the EM-PIC code OSIRIS \cite{OSIRIS}, as well as UCLA PIC framework \cite{UPIC}, and good agreement is obtained. 

Moreover, by plotting the intersection of the EM  and beam modes in $(k_1,k_2,w_r)$ space the numerical instability patterns can be conveniently predicted.   Maxwell Solvers, such as spectral solvers, that have EM waves with phase velocities greater than the plasma drifting velocity are therefore free of any instability due to the main beam mode. We showed that the use of a spectral solver does indeed eliminate any instability from the main beam mode,  leading to instability predominantly from coupling between the EM wave and the lowest order aliased beam modes. This coupling  exists only at high $\vert\vec{k}\vert$ area at the edge of the lowest order Brillouin zone. We showed that a low pass filter with a hard cutoff can eliminate these modes  without effecting lower $\vert\vec k\vert$ modes that are physically important in a properly resolved simulation.  

In addition, by using the fact that the modes with highest growth rate are found near the intersection of the beam modes and EM modes, we derived an asymptotic expression for the growth rate that can be useful to study the growth rates with various smoothing functions and different cell sizes. The asymptotic expression speeds up the calculation of instability growth rate and makes the investigation of instability pattern and growth rate in 3D feasible. By conducting parameters scan using the asymptotic expression for the Yee, and Karkkainen solver in 2D and 3D, we confirmed the ``magic time step'' that minimize the growth rate, as reported in \cite{VayJCP2011}. We found the ratio of the magic time step over the grid size along the drifting direction, $\Delta t_m/\Delta x_1$ is determined by the field interpolation scheme used in the simulation, yet stays constant for different $\Delta x_1$. These observations agree well with the simulation results. 

This paper reports on efforts to understand and mitigate the instability in cases in which there is only a drifting plasma. Areas for  future work include trying to optimize the particle order, field smoother, and field solver for mitigating this numerical instability in LWFA simulation, understanding how the various solvers and use of specific time steps effect the numerical dispersion relation for light waves, and for studying if the use of the Yee mesh together with a spectral solver also leads to an optimal time step. In addition, future work should include exploring the tradeoffs in accuracy, computational speed, and parallel scalability for the different choices.

This work was supported by  DOE awards DE-FC02-07ER41500,  DE-SC0008491,  DE-FG02-92ER40727,  and DE-SC0008316,  and by NSF grants NSF PHY-0904039 and NSF PHY-0936266. Simulations were carried out on the UCLA Hoffman 2 Cluster.


\appendix

\section{Interpolation dyadic and finite difference operator}
\label{sect:app:s}
In this appendix we will write out the explicit expressions for the interpolation dyadics $\overleftrightarrow{S}$ for the fields and the currents. For a momentum conserving scheme, in 3D the interpolation dyadic for the EM field after the Fourier transform can be expressed as:
\begin{align}
S_{E1}&=s_{l,1}s_{l,2}s_{l,3}\eta_1\quad S_{E2}=s_{l,1}s_{l,2}s_{l,3}\eta_2\quad S_{E3}=s_{l,1}s_{l,2}s_{l,3}\eta_3\nonumber\\
S_{B1}&=\cos(\omega\Delta t/2)s_{l,1}s_{l,2}s_{l,3}\eta_2\eta_3\quad S_{B2}=\cos(\omega\Delta t/2)s_{l,1}s_{l,2}s_{l,3}\eta_1\eta_3\nonumber\\ S_{B3}&=\cos(\omega\Delta t/2)s_{l,1}s_{l,2}s_{l,3}\eta_1\eta_2
\end{align}
where
\begin{align}
s_{l,i}=\biggl(\frac{\sin(k_i\Delta x_i/2)}{k_i\Delta x_i/2}\biggr)^{l+1}
\end{align}
and $\eta_i=\zeta^{\nu_i}$, $\zeta=-1$ when the EM field has half-grid offset in the $\hat{i}$ direction, and $\zeta=1$ when it is defined at grid point (this was the term missing in our earlier version). $l$ refers to the order ($l=1$ is area weighting or linear interpolation for the charge).

For an energy conserving scheme, we have 
\begin{align}
S_{E1}&=s_{l-1,1}s_{l,2}s_{l,3}\eta_1\quad
S_{E2}=s_{l,1}s_{l-1,2}s_{l,3}\eta_2\quad
S_{E3}=s_{l,1}s_{l,2}s_{l-1,3}\eta_3\nonumber\\
S_{B1}&=\cos(\omega\Delta t/2)s_{l,1}s_{l-1,2}s_{l-1,3}\eta_2\eta_3\quad
S_{B2}=\cos(\omega\Delta t/2)s_{l-1,1}s_{l,2}s_{l-1,3}\eta_1\eta_3\quad\nonumber\\
S_{B3}&=\cos(\omega\Delta t/2)s_{l-1,1}s_{l-1,2}s_{l,3}\eta_1\eta_2
\end{align}
For the rigorous charge conserving scheme (as is used in OSIRIS), the  current interpolation dyadic is approximately:
\begin{align}
S_{j1}=s_{l-1,1}s_{l,2}s_{l,3}\eta_1\quad
S_{j2}=s_{l,1}s_{l-1,2}s_{l,3}\eta_2\quad
S_{j3}=s_{l,1}s_{l,2}s_{l-1,3}\eta_3
\end{align}
We note that the expressions for $S_j$ are strictly valid for the charge conserving scheme in the limit of $\Delta t \rightarrow 0$. Meanwhile, when the current is directly deposited (as is done in the UPIC framework),  the current interpolation functions are, 
\begin{align}
S_{j1}=s_{l,1}s_{l,2}s_{l,3}\eta_1\quad
S_{j2}=s_{l,1}s_{l,2}s_{l,3}\eta_2\quad
S_{j3}=s_{l,1}s_{l,2}s_{l,3}\eta_3
\end{align}

The space finite difference operator for the Yee solver is:
\begin{align}
[k]_i = \frac{\sin(k_i\Delta x_i/2)}{\Delta x_i/2}
\end{align}
and is the same for electric and magnetic field. The space finite difference operator for the Karkkainen solver is the same as Yee solver for the magnetic field, as for the electric field, we used
\begin{align}
[k]_{Ei} = c_i\frac{\sin(k_i\Delta x_i/2)}{\Delta x_i/2}
\end{align}
where
\begin{align}
c_1 &= \theta_1 + 2\theta_2\{\cos(k_2\Delta x_2)+\cos(k_3\Delta x_3)\}+4\theta_3\cos(k_2\Delta x_2)\cos(k_3\Delta x_3)\nonumber\\
c_2 &= \theta_1 + 2\theta_2\{\cos(k_3\Delta x_3)+\cos(k_1\Delta x_1)\}+4\theta_3\cos(k_3\Delta x_3)\cos(k_1\Delta x_1)\nonumber\\
c_3 &= \theta_1 + 2\theta_2\{\cos(k_1\Delta x_1)+\cos(k_2\Delta x_2)\}+4\theta_3\cos(k_1\Delta x_1)\cos(k_2\Delta x_2)
\end{align}
and
\begin{align}
\theta_1 = 7/12\qquad \theta_2 = 1/12 \qquad \theta_3=1/48
\end{align}
are the tunable parameters for the Karkkainen solver \cite{kark}. The space finite difference operator for the spectral solver is
\begin{align}
[k]_i=k_i
\end{align}

The time finite difference operator for the Yee, Karkkainen, and spectral solver are the same
\begin{align}
[\omega]=\frac{\sin(\omega\Delta t/2)}{\Delta t/2}
\end{align}


\end{document}